\documentclass[10pt, conference, letterpaper]{IEEEtran}
\usepackage{xcolor}
\IEEEoverridecommandlockouts
\usepackage{amsmath,amssymb,amsfonts}
\usepackage{gensymb}
\usepackage{array}
\usepackage{algorithmic}
\usepackage{graphicx}
\usepackage{textcomp}
\usepackage{xcolor}
\usepackage{xspace}
\usepackage{enumitem}
 \usepackage{url}
\usepackage{breakurl}
\usepackage{caption}
\usepackage{subcaption}
\usepackage{multirow}
\usepackage[normalem]{ulem}
\usepackage[numbers, sort&compress]{natbib}
\def\BibTeX{{\rm B\kern-.05em{\sc i\kern-.025em b}\kern-.08em
    T\kern-.1667em\lower.7ex\hbox{E}\kern-.125emX}}
\graphicspath{ {./images/} }

\newcommand{\PreserveBackslash}[1]{\let\temp=\\#1\let\\=\temp}
\newcolumntype{C}[1]{>{\PreserveBackslash\centering}p{#1}}
\newcolumntype{R}[1]{>{\PreserveBackslash\raggedleft}p{#1}}
\newcolumntype{L}[1]{>{\PreserveBackslash\raggedright}p{#1}}

\newcommand{\ie}{\textit{i.e.,}\xspace}

\begin{document}


\title{A Comprehensive Real-World Evaluation of 5G Improvements over 4G in Low- and Mid-Bands 
\thanks{\hrule \vspace{4pt} This research is supported in part by the National Science Foundation under grant number CNS-2128489, 2132700, 2220286, 2220292, 2226437, and 2229387.}
}

\author{\IEEEauthorblockN{
Muhammad Iqbal Rochman\IEEEauthorrefmark{1},
Wei Ye\IEEEauthorrefmark{2},
Zhi-Li Zhang\IEEEauthorrefmark{2},
and Monisha Ghosh\IEEEauthorrefmark{3}}
\IEEEauthorblockA{\IEEEauthorrefmark{1}University of Chicago, 
\IEEEauthorrefmark{2}University of Minnesota Twin Cities,
\IEEEauthorrefmark{3}University of Notre Dame.\\ 
Email: \IEEEauthorrefmark{1}muhiqbalcr@uchicago.edu, \IEEEauthorrefmark{2}\{ye000094,zhzhang\}@umn.edu, \IEEEauthorrefmark{3}mghosh3@nd.edu}}


\maketitle

\begin{abstract}

As discussions around 6G begin, it is important to carefully quantify the spectral efficiency gains actually realized by deployed 5G networks as compared to 4G through various enhancements such as higher modulation, beamforming, and MIMO.
This will inform the design of future cellular systems, especially in the mid-bands, which provide a good balance between bandwidth and propagation.
Similar to 4G, 5G also utilizes low-band ($<$1 GHz) and mid-band spectrum (1 to 6 GHz), and hence comparing the performance of 4G and 5G in these bands will provide insights into how further performance improvements can be attained.
In this work, we address a crucial question: is the performance boost in 5G compared to 4G primarily a result of increased bandwidth, or do the other enhancements play significant roles, and if so, under what circumstances? There is extremely limited academic research that addresses this important question.
Hence, we conduct city-wide measurements of 4G and 5G cellular networks deployed in low- and mid-bands in Chicago and Minneapolis, and carefully analyze the performance to quantify the contributions of different aspects of 5G advancements to its improved throughput performance.
Our analyses show that (i) compared to 4G, the throughput improvement in 5G today is mainly influenced by the wider channel bandwidth, both from single channels and channel aggregation, (ii) in addition to wider channels, improved 5G throughput requires better signal conditions, which can be delivered by denser deployment and/or use of beamforming in mid-bands, (iii) the channel rank in real-world environments rarely supports the full 4 layers of 4x4 MIMO and (iv) advanced features such as MU-MIMO and higher order modulation such as 1024-QAM have yet to be widely deployed.
These observations and conclusions lead one to consider designing the next generation of cellular systems to have wider channels, perhaps with improved channel aggregation, a deployment architecture that is dense and uses more beams, thus ensuring uniformly better signal strength over the coverage area and no more than 4 MIMO layers per user.


\end{abstract}


\section{Introduction}\label{sec:intro}

5G New Radio (NR) networks are widely recognized as the cornerstone for emerging next-generation mobile applications such as AR/VR, remote driving, and cloud gaming. Our recent measurement studies~\cite{carpenter2023multi} reveal that with good coverage, current commercial 5G networks can achieve an average downlink throughput of 700 Mbps (with peak rates reaching $\sim$1 Gbps) even while driving in urban environments, far surpassing the performance of 4G Long-Term Evolution (LTE) networks. This substantial improvement may be attributed to the adoption of advanced wireless communication technologies, including but not limited to massive MIMO~{\cite{bogale2016massive}}, beamforming~{\cite{uwaechia2020comprehensive}}, higher modulation~{\cite{3gpp17-38214}}, and increased bandwidth. The mid-band spectrum (1--6 GHz) in particular is deemed crucial for the success of 5G, as it balances coverage with bandwidth ~{\cite{midband}}.
However, some advanced 5G technologies may still be in the product development phase or skipped altogether due to cost considerations.
Additionally, due to the phased deployment of commercial 5G, earlier research does not have results on the latest deployment of 5G in the BRS-band (2.5--2.7 GHz) and C-band (3.7--4.2 GHz). It is essential to quantify the spectral utilization of 5G to determine if additional spectrum should be allocated to mobile broadband.

\begin{table}[t]
\centering
\caption{Statistics of 4G/5G dataset.}
\vspace{-.5em}
\resizebox{\columnwidth}{!}{
\begin{tabular}{|c|C{4cm}|} \hline 
    \textbf{Mobile Operators} & AT\&T, T-Mobile, Verizon\\ \hline 
    \textbf{Radio Technologies} & 4G, 5G\\ \hline 
    \textbf{Measurement Venues} & Minneapolis, Chicago\\ \hline 
    \textbf{Cumulative Data Traces} & 1200+km; Around 14 hours\\ \hline
    \textbf{XCAL Key Perf. Indicators} & PCI-Beam idx; Freq.; SCS; RSRP; RSRQ; CQI; RI; BLER; MCS; \#RBs; \#MIMO layers; MIMO modes; PHY-layer throughput; \\ \hline
 \end{tabular}
}
\label{tab:data-statistics}
\vspace{-2em}
\end{table}

\begin{figure*}[t]
    \centering
    \includegraphics[width=\textwidth, height=2.75in]{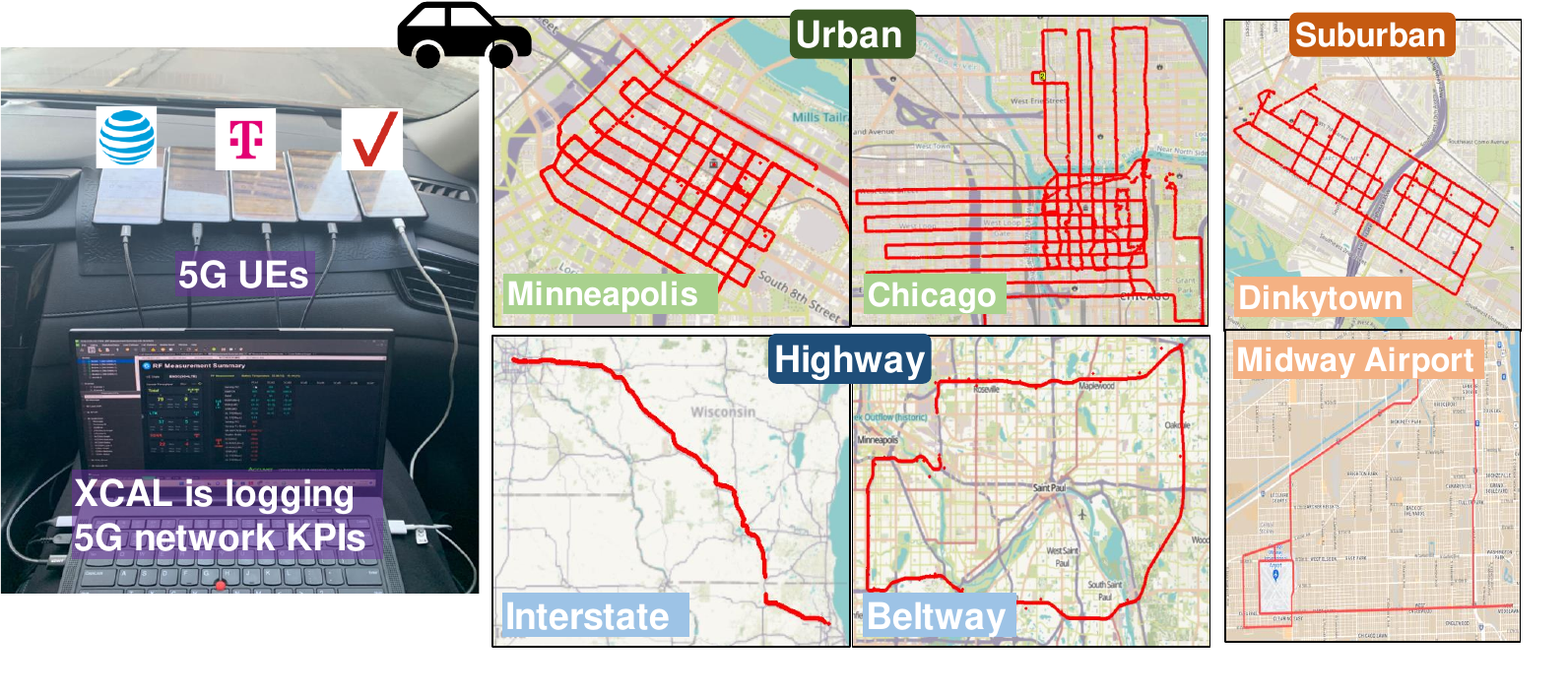}
    \vspace{-3em}
    \caption{Setup of the 5G probes and driving routes of our measurements.}
    \label{fig:data-collection-map}
    \vspace{-1.5em}
\end{figure*}

Hence, in this paper we focus on the physical layer of 4G and 5G networks, exploring deployed radio channel characteristics and their impact on throughput performance, with emphasis on the low-band ($<$1 GHz) and mid-band channels, particularly the recent deployments of NR in BRS~\cite{tmb-2.5GHz} and C-band~\cite{att-3.7GHz,vzw-3.7GHz}. Specifically, we wish to address the following questions: What are the similarities and differences in the channel resources and radio technologies currently used in 4G and 5G networks? More importantly, is the observed 5G performance boost primarily a result of increased bandwidth, or do other novel enhancements play significant roles? And can we quantify their relationship and contribution to throughput? We believe that addressing these queries can prompt a reevaluation of mobile network evolution, offering valuable insights for the design and deployment of future 6G networks.
 
To this end, we carried out extensive in-field measurements in two major metropolitan areas in the US, Minneapolis and Chicago, along with the highways connecting them, as shown in Fig.~\ref{fig:data-collection-map}.
We employ commercial smartphones as user equipment (UE) and utilize a professional network analyzer tool to log all network events and key performance indicators (KPIs).
Our data trajectory spans a driving distance exceeding 1200 km, encompassing both 4G and 5G networks of three major US operators—namely, AT\&T (ATT), T-Mobile (TMO), and Verizon (VZW) as shown in Table~\ref{tab:data-statistics} (see \textbf{\S\ref{sec:methodology}} for details).
To the best of our knowledge, our work is one of the first to conduct a comprehensive city-wide study spanning two cities, aimed at assessing the enhancements offered by 5G over 4G, with a particular focus on the low- and mid-bands.

With these datasets, we conduct an in-depth data analysis and summarize our conclusions as follows:

\noindent\textbf{(1) Comparison of throughput performance between NR and LTE (\S\ref{sec:tput-comp}:}
We perform a comparative analysis of throughput over the representative low- and mid-band channels in NR and LTE. As expected, the NR mid-band channels perform exceedingly well in downlink and uplink compared to other NR and LTE channels. However, we further normalize the throughput and demonstrate that the higher channel bandwidth of mid-band 5G is a major factor in the increased throughput, rather than new 5G features. We observe similar results for uplink throughput and hence do not specifically present those results.

\noindent\textbf{(2) Impact of deployment density and beamforming on downlink throughput (\S\ref{sec:deployment}):} We compare the density of base stations (BSs) deployed by the three carriers and beamforming modes used and note that denser deployments along with the use of more beams in the mid-band lead to improved overall signal strength and hence to improved spectral efficiency.

\noindent\textbf{(3) Analysis of the contribution of various 4G and 5G signal parameters to normalized downlink throughput (\S\ref{sec:contrib-analysis}):}
The contribution of various signal parameters such as Reference Signal Received Power (RSRP), Modulation Coding Scheme (MCS), and Channel Quality Indicator (CQI) to the normalized throughput is further analyzed. We demonstrate that these parameters in NR perform similarly to their LTE counterparts. We also note the absence of 1024-QAM modulation, introduced in the latest 3GPP Release 17.

\noindent\textbf{(4) Comparison of Multiple-Input Multiple-Output (MIMO) performance in NR and LTE (\S\ref{sec:mimo-comp}):}
A comparative analysis of the Rank Indicator (RI) value shows a marginal increase in NR MIMO performance compared to LTE in terms of the actual channel rank and number of layers that the physical channel can support. We also note that Multi-User MIMO (MU-MIMO) has not yet been deployed in the cities we studied.

\noindent\textbf{(5) Comparison of latency performance between NR-SA (Standalone), NR-NSA (Non-Standalone), and LTE networks in low- and mid-bands for T-Mobile (\S\ref{sec:tput-comp}, \S\ref{sec:lat-comp}):}
Since T-Mobile is the only operator with widely deployed NR-SA, we compared the latency performance of NR-SA, NR-NSA, and LTE. Mid-band NR-SA demonstrates the best latency performance due to the combination of lower signaling overhead (compared to NR-NSA) and denser mid-band deployments.


\section{Background and Related Work}
\label{sec:related_work}

The current 5G specification incorporates numerous advanced technologies to enhance efficiency, including high-order quadrature amplitude modulation (QAM), massive MIMO, and beamforming.
High-order QAM enables transmission of more information bits within a symbol interval, and the 3rd Generation Partnership Project (3GPP) introduced 1024-QAM in Release 17 of the 5G specification~\cite{3gpp17-38214}.
Beamforming optimizes signal direction, concentrating energy on specific users for improved coverage and capacity. NR beamforming is particularly important for mmWave channels~\cite{uwaechia2020comprehensive,li2020beam}, and our prior work~\cite{narayanan2022comparative} studied the deployment of mmWave 5G beamforming in Chicago and the correlation of the number of beam indices to beam width and beam handover performance. Beamforming can also be implemented in low- and mid-band channels, with the standard specifying up to 8 beams in the mid-band TDD spectrum. These beams are identified with their unique Synchronization Signal Block (SSB) index.
Massive MIMO, or MU-MIMO, utilizes beamforming to achieve a higher degree of spatial multiplexing between multiple users simultaneously, but it is known to suffer from a fundamental problem called ``pilot contamination''~\cite{elijah2015comprehensive,jose2011pilot,bogale2016massive}. In 5G, this problem is alleviated using channel estimation from channel state information-reference signal (CSI-RS) in downlink for frequency division duplexing (FDD) systems, and sounding reference signal (SRS) in the uplink for time division duplexing (TDD) systems~\cite{lopez2022survey, 3gpp17-38214}.

While prior 4G mobile networks mainly operated on the low ($<$1 GHz) and mid (1--6 GHz) frequency bands (collectively referred to as frequency range 1, or FR1~\cite{3gpp17-38101-1}), initial 5G deployments in the US focused on the mmWave bands ($>$24 GHz), also known as FR2.
The mmWave bands can provide a throughput performance of over 1 Gbps, but performance is limited by propagation loss, body loss, foliage, and thermal effects as shown in our prior analyses~\cite{narayanan2022comparative,rochman2023comprehensive}.
The low-band, conversely, offers better coverage but with relatively lower bandwidth and hence throughput.
Therefore, the mid-band which strikes a balance between coverage and performance has become a focal point for current 5G deployments.
For instance, TMO is actively advancing its deployment in the BRS band~{\cite{tmb-2.5GHz}}, while ATT and VZW are strategically deploying in the C-band ~{\cite{att-3.7GHz,vzw-3.7GHz}}. Simultaneously, VZW also uses the CBRS band (3.55--3.7 GHz) in its 4G network, using both Tier 2 Priority Access License (PAL) and Tier 3 General Authorized Access (GAA) modes~{\cite{vzw-CBRS}}. Furthermore, ATT and VZW have further extended their 4G networks to the unlicensed 5 GHz spectrum through LAA~{\cite{vzw-att-LAA}}.


The novel features specified in 5G can theoretically improve throughput performance since network throughput is a function of channel bandwidth, modulation, code rate, and number of MIMO layers~\cite{3gpp17-38306}.
However, the reality is much more complex and presents numerous challenges. For example, receiver and transmitter hardware may limit the adoption of those new technologies; the regulator will constrain the transmission power in some locations; radio channels located at higher frequencies will further suffer from radio interference and fading; and poor channel quality will significantly reduce the efficiency of higher-order QAM due to increased transmission error.
Prior real-world deployment studies of commercial 5G did not delve into the fundamental reasons for performance enhancement or contrast the improvements to its 4G counterpart~{\cite{carpenter2023multi, fezeu2023mid, narayanan2021variegated, ghoshal2023performance, rochman2023comprehensive}}. 
Authors of \cite{liu2023close} have highlighted missed throughput potential in deployed 5G and offered possible solutions, but the work did not compare 5G performance with 4G and the analysis was performed at the link layer.
We believe that the work presented in this paper is the first to comprehensively analyze the contribution of different physical layer techniques on the throughput performance of 5G as compared to 4G.


\section{Measurement Settings and Methodology}\label{sec:methodology}

The data analyzed in this paper was collected over three measurement campaigns. We conducted an initial campaign in Chicago during December 2022, followed by a campaign in Minneapolis during April-May and November 2023, with both focusing on the downlink and uplink throughput performance of the three major US operators (ATT, TMO, and VZW). Additionally, a latency-focused campaign specifically targeting the T-Mobile network was conducted in Minneapolis in March 2024.
To cover a large area, we conducted data collection while driving. Fig.~\ref{fig:data-collection-map} shows the setup of 5G probes and the driving routes: in Chicago, we measured in the downtown Loop area, Midway airport, and the interstate highway; while in Minneapolis, we surveyed the Downtown, Dinkytown, and the beltway.
The statistics of the data collected are summarized in Table~\ref{tab:data-statistics}.

All the data were captured using Samsung Galaxy S22+ (Android 12), which can receive 5G signals in the low-band, mid-band, and mmWave channels. At the time, S22+ was the only phone capable of capturing the TMO network's inter-band 5G Carrier Aggregation (CA). These state-of-the-art devices allowed us to measure the network's best possible performance.
Three S22+ phones were used as user equipment (UE), each equipped with ATT, TMO, and VZW SIMs. All SIMs have unlimited data plans with no throttling of data rates.
The S22+ phones were connected to a Lenovo ThinkPad X1 Carbon laptop running Accuver XCAL~\cite{xcal}. The XCAL application collects various 4G and 5G signal parameters by establishing a low-level interface to the modem chipset. 
To account for differences in parameter sampling intervals, the application processes the data over one-second periods, averaging numerical values and determining the most frequent (mode) for discrete values. Table~\ref{tab:data-statistics} also summarizes the parameters (key performance indicators) collected by XCAL for our analysis.
XCAL is also capable of actively creating traffic using iperf~\cite{iperf} tools: we generate full-buffer downlink (iperf server to UE) and uplink (UE to iperf server) transmission to cloud servers in Chicago and Minneapolis, whichever is closest to the measurement location.

\begin{table}[t]
\centering
\caption{Highlight of features in 3GPP Rel-16 and Rel-17 compared to observed 4G and 5G in our dataset.}
\label{tab:4g5g-features}
\vspace{-.5em}
\resizebox{\columnwidth}{!}{
\begin{tabular}{|C{2cm}|c|c|c|c|} \hline 
\textbf{Parameters}      & \textbf{Observed 4G} & \textbf{Observed 5G} & \textbf{Rel-16 5G} & \textbf{Rel-17 5G} \\ \hline 
\textbf{Max. Modulation} & 256-QAM              & 256-QAM              & 256-QAM            & 1024-QAM           \\ \hline 
\textbf{Max. MIMO Layer} & 4                    & 4                    & 8                  & 8                  \\ \hline 
\textbf{Max. Ch. BW (excl. mmWave)}  & 20 MHz               & 100 MHz              & 100* MHz        & 100* MHz            \\ \hline
\textbf{Max. \#CA}       & 6                    & 4                    & 16                 & 16                 \\ \hline
\end{tabular}
}
\footnotesize{*mmWave channels can be up to 400 MHz wide}
\vspace{-1em}
\end{table}

\begin{table}[t]
\caption{NR and LTE Bands Information}
\vspace{-.5em}
\label{tab:nr-lte-bands}
\centering
\resizebox{.95\linewidth}{!}{
\begin{tabular}{|C{.075\textwidth}|C{.05\textwidth}|C{.095\textwidth}|C{.04\textwidth}|C{.05\textwidth}|}
\hline
\textbf{Operator-Band} & \textbf{Duplex Mode} & \textbf{DL Band Freq. (MHz)} & \textbf{SCS (kHz)} & \textbf{BW (MHz)} \\ \hline \hline
\multicolumn{5}{|c|}{\textbf{Representative Bands}} \\ \hline
ATT-n5                 & FDD                  & 850                          & 15                & 10                \\
ATT-n77                & TDD                  & 3700                         & 30                & 40                \\ \hline
TMO-n41                & TDD                  & 2500                         & 30                & 40,100            \\
TMO-n71                & FDD                  & 600                          & 15                & 20                \\ \hline
VZW-n5                 & FDD                  & 850                          & 15                & 10                \\
VZW-n77                & TDD                  & 3700                         & 30                & 60                \\ \hline
ATT-b2                 & FDD                  & 1900                         & 15                & 20                \\
ATT-b12                & FDD                  & 700                          & 15                & 10                \\
ATT-b46                & TDD                  & 5200                         & 15                & 20                \\ \hline
TMO-b12                & FDD                  & 700                          & 15                & 5                 \\
TMO-b41                & FDD                  & 2500                         & 15                & 20                \\
TMO-b66                & FDD                  & 2100                         & 15                & 20                \\ \hline
VZW-b13                & FDD                  & 700                          & 15                & 10                \\
VZW-b48                & TDD                  & 3500                         & 15                & 10,20             \\
VZW-b66                & FDD                  & 2100                         & 15                & 20                \\ \hline \hline
\multicolumn{5}{|c|}{\textbf{Other Bands}} \\ \hline
ATT-n66                & FDD                  & 2100                         & 15                & 5                 \\ \hline
TMO-n25                & FDD                  & 1900                         & 15                & 20                \\ \hline
ATT-b14                & FDD                  & 700                          & 15                & 10                \\
ATT-b29                & SDL                  & 700                          & 15                & 5                 \\
ATT-b30                & FDD                  & 2300                         & 15                & 5,10              \\
ATT-b66                & FDD                  & 2100                         & 15                & 5,10,15           \\ \hline
TMO-b2                 & FDD                  & 1900                         & 15                & 10                \\
TMO-b4                 & FDD                  & 2100                         & 15                & 20                \\
TMO-b25                & FDD                  & 1900                         & 15                & 10                \\
TMO-b46                & TDD                  & 5200                         & 15                & 20                \\
TMO-b71                & FDD                  & 600                          & 15                & 5                 \\ \hline
VZW-b2                 & FDD                  & 1900                         & 15                & 10                \\
VZW-b4                 & FDD                  & 2100                         & 15                & 20                \\
VZW-b5                 & FDD                  & 850                          & 15                & 10                \\
VZW-b46                & TDD                  & 5200                         & 15                & 20                \\ \hline
\end{tabular}
}
\vspace{-1.5em}
\end{table}

\begin{figure*}[t]
    \begin{subfigure}{.32\textwidth}
    \includegraphics[width=\linewidth]{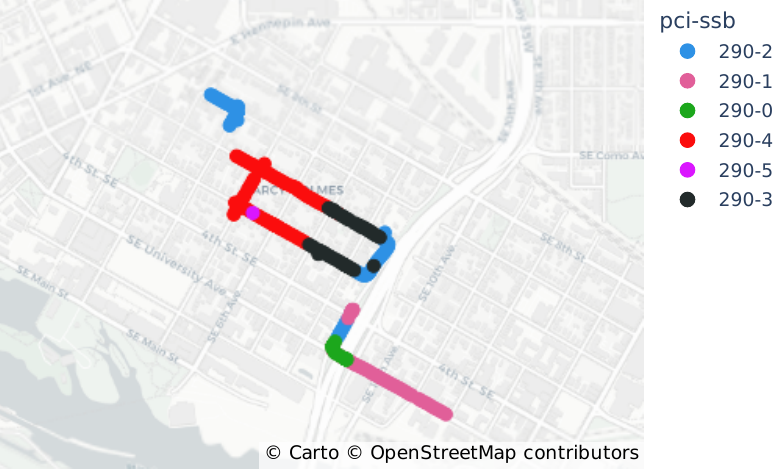}
    \vspace{-1.5em}
    \caption{ATT-n77 PCI 290}
    \label{fig:map-pci-ssb-nr-att}
    \end{subfigure}
    \hfill
    \begin{subfigure}{.32\textwidth}
    \includegraphics[width=\linewidth]{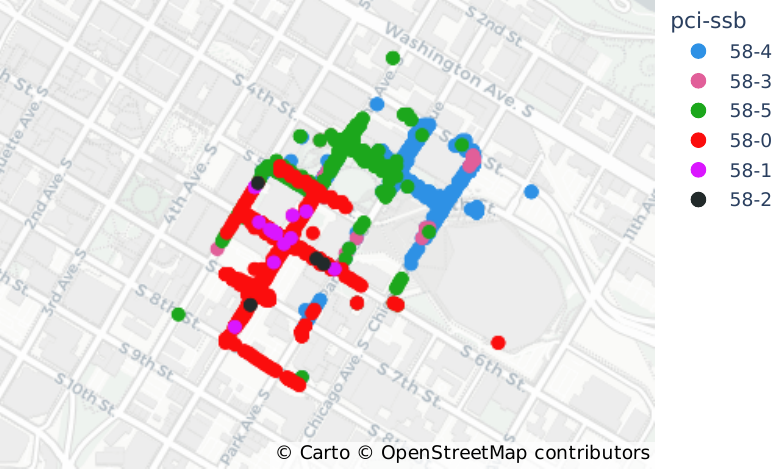}
    \vspace{-1.5em}
    \caption{TMO-n41 PCI 58}
    \label{fig:map-pci-ssb-nr-tmo}
    \end{subfigure}
    \hfill
    \begin{subfigure}{.32\textwidth}
    \includegraphics[width=\linewidth]{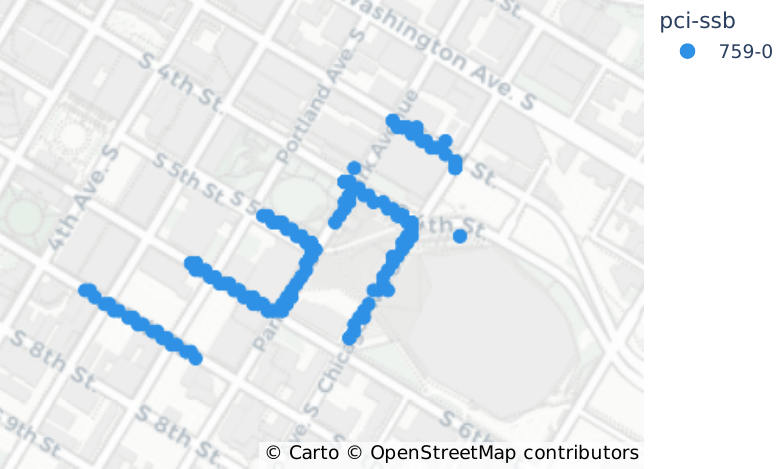}
    \vspace{-1.5em}
    \caption{VZW-n77 PCI 759}
    \label{fig:map-pci-ssb-nr-vzw}
    \end{subfigure}
    \vspace{-.5em}
    \caption{PCI-SSB index maps of mid-band channels.}
    \label{fig:map-pci-ssb-nr}
    \vspace{-1.5em}
\end{figure*}

\section{Measurement Results and Analyses}\label{sec:results}

\begin{table}[t]
\caption{Comparison of Rep. Bands' Deployment}
\vspace{-.5em}
\label{tab:nr-lte-bands-reps}
\centering
\footnotesize{(Bands in bold are mid-bands.)\vspace{.5em}}
\resizebox{.95\linewidth}{!}{
\begin{tabular}{|C{.075\textwidth}|C{.05\textwidth}|C{.05\textwidth}|C{.05\textwidth}|C{.06\textwidth}|}
\hline
\textbf{Operator-Band} & $\mathbf{n}$ \textbf{data} & \textbf{\% data} & \textbf {\#SSB Inds.} & \textbf{\#unique PCI}\\ \hline \hline
\textbf{ATT-n77}       & 8380        & 34                & 1,6*                 & 152                  \\
ATT-n5                 & 14444       & 58                & 1                    & 217                  \\ \hline
\textbf{TMO-n41}       & 56606       & 90                & 6                    & 464                  \\
TMO-n71                & 1981        & 3                 & 1                    & 60                  \\ \hline
\textbf{VZW-n77}       & 13049       & 96                & 1                    & 147                  \\
VZW-n5                 & 541         & 3                 & 1                    & 28                  \\ \hline
\textbf{ATT-b2}        & 31480       & 28                & N/A                  & 378                  \\
\textbf{ATT-b46}       & 7090        & 6                 & N/A                  & 77                  \\
ATT-b12                & 12690       & 11                & N/A                  & 262                 \\ \hline
\textbf{TMO-b66}       & 19334       & 39                & N/A                  & 330                  \\
\textbf{TMO-b41}       & 7400        & 15                & N/A                  & 54                  \\
TMO-b12                & 2645        & 5                 & N/A                  & 83                  \\ \hline
\textbf{VZW-b66}       & 31742       & 36                & N/A                  & 379                  \\
\textbf{VZW-b48}       & 8762        & 8                 & N/A                  & 141                  \\
VZW-b13                & 16906       & 19                & N/A                  & 255                  \\ \hline
\end{tabular}
}
\\ \vspace{.2em}
\footnotesize{*ATT-n77 has 6 SSB indices in Minneapolis, but only 1 in Chicago.}
\vspace{-1.5em}
\end{table}

\subsection{Overview of the Observed 4G and 5G Deployments}

Table~\ref{tab:4g5g-features} compares the 4G and 5G features we observed to the 3GPP specifications in Release 16 and 17. We believe that most deployments today are at most Release 16.
Up to 256-QAM is observed in both LTE and NR networks,  but not 1024-QAM. 
We also did not observe improvements in the number of MIMO layers/streams for low- and mid-band 5G over the 4G counterparts, even though 3GPP Rel-16 supports up to 8 layers.
On the other hand, there are improvements in maximum channel bandwidth, as new spectrum has been released.
This is reflected in the reduced number of aggregated channels: as bandwidth increases, there is less need to increase CA in 5G.

NR has two modes of network deployments: SA which only utilizes the 5G channels and network stack, and NSA which utilizes a combination of 4G and 5G channels and stacks with 4G used as the primary carrier.
In all campaigns, ATT and VZW deployed only NSA where NR and LTE channels are aggregated, but we did not observe two NR low- or mid-band channels aggregated in their network. On the other hand, TMO deployed both SA and NSA modes, with the SA mode aggregating up to 3 NR channels for a total bandwidth of 160 MHz.
Table~\ref{tab:nr-lte-bands} shows the summary of captured NR and LTE bands/channels in the campaign. NR channels have the prefix "n" and LTE channels have the prefix "b" in the table. All three operators have deployed NR in low- and mid-bands. Notably, ATT-n66 and TMO-n25 are the newest bands detected only in our April-May 2023 campaign and afterward. AT\&T and Verizon have also deployed NR mmWave bands (n260 and n261), but they are out of scope for this analysis.
Among the NR low- and mid-band channels, all FDD channels are deployed with 15 kHz sub-carrier spacing (SCS) and lower bandwidth (\ie ATT-n5, ATT-n66, TMO-n25, TMO-n71, VZW-n5), while all TDD bands are deployed with 30 kHz SCS and higher bandwidth (\ie ATT-n77, TMO-n41, VZW-n77). These deployments suggest that the NR FDD bands are positioned as the ``support'' bands since the lower bandwidth and frequency result in lower throughput but greater coverage. It should be noted that TMO's mid-band deployment in 2.5 GHz has a 3.4 dB advantage over ATT and VZW in 3.7 GHz: this will be seen in performance results presented later.
Most of the deployed LTE channels are FDD, except for the newer b46 (LAA) and b48 (CBRS) which are TDD, and b29 which is a supplementary downlink (SDL) band. The LTE bands are similarly used as the ``support'' bands to the NR TDD  bands when aggregated in the NSA deployments.


\subsection{Comparison of Low- and Mid-band Deployment}\label{sec:deployment}

Preliminary analysis revealed negligible differences between the datasets collected in Chicago and Minneapolis. Consequently, we combined them for the analyses presented in this paper. For brevity, we selected a representative low-band and mid-band channel for each operator with substantial data points, regardless of their primary or secondary cell/channel designation.
All mid-band channels are deployed with higher bandwidth compared to low-band channels. 
Table~\ref{tab:nr-lte-bands-reps} shows the number of data points and deployment parameters of the selected representative bands. Due to PCI reuse, the number of unique Physical Cell Identifiers (PCIs) is calculated separately between Chicago and Minneapolis.
For ATT NR, n77 (mid-band) and n5 (low-band) are selected. There are a larger number of data points and unique PCIs on n5, indicating a denser deployment in the low-band: this is in contrast to TMO and VZW which have a higher number of data points and unique PCIs in mid-bands (n41 and n77, respectively) compared to low-bands (n71 and n5, respectively). Moreover, TMO-n41 is very densely deployed with 462 unique PCIs compared to the NR channels from ATT and VZW. As we shall see later, the density of the TMO NR deployment and the lower NR mid-band frequency (2.5 GHz) significantly impact overall superior signal strength, spectral efficiency and throughput compared to ATT and VZW.

Similar to the NR bands, we observe a higher number of data points and unique PCIs on mid-band LTE channels compared to low-band. For ATT, b66 has the largest number of data points for mid-band. However, we chose the second-largest b2 for its wider bandwidth. For TMO, b12 is the only low-band channel beside b71, indicating the very sparse deployment of low-band LTE in TMO: b12 and b71 channels account for 10\% of the total LTE data.
The proportion of mid-band data in LTE and NR indicates that TMO and VZW have been focusing on mid-band deployments.
Lastly, we selected ATT-b46 and VZW-b48 as representatives of the unlicensed/shared mid-band 4G channels. Additionally, TMO-b41 is chosen for comparison with TMO-n41. These bands exhibited fewer data points and unique PCI compared to other channels. This suggests localized deployments using small cells: at least b46 and b48 are limited in the transmit power.

\vspace{.5em}\noindent\textit{\textbf{Comparison of low- and mid-band NR beam deployment:}}
Table~\ref{tab:nr-lte-bands-reps} also shows the number of SSB Indices which denotes the number of beams available for each NR channel.
We observe only index 0 (\ie single beam) per PCI for ATT channels in the Chicago campaign. However, we observe 6 SSB indices for ATT-n77 in the later Minneapolis campaign. Fig.~\ref{fig:map-pci-ssb-nr-att} shows the coverage of the various SSB Indices for ATT-n77, PCI 290, in Dinkytown.
Similarly, Fig.~\ref{fig:map-pci-ssb-nr-tmo} depicts SSB Indices from 0 to 5 for TMO-n41, PCI 59, in downtown Minneapolis. All low-band NR channels, \ie n5 and n71, only use one SSB index, and the number of SSB indices did not change between the two campaigns. 
Unlike TMO and ATT, VZW always uses a single SSB index per PCI for all of its NR channels in both cities. Fig.~\ref{fig:map-pci-ssb-nr-vzw} illustrates the coverage of one of VZW's n77 channels in downtown Minneapolis. Since more SSBs/PCI means more beams and hence beamforming gain, VZW with only one beam/PCI suffers from lower signal strength overall and poorer spectral efficiency, as will be shown later.


\begin{figure}[t]
    \vspace{-.5em}
    \includegraphics[width=\linewidth]{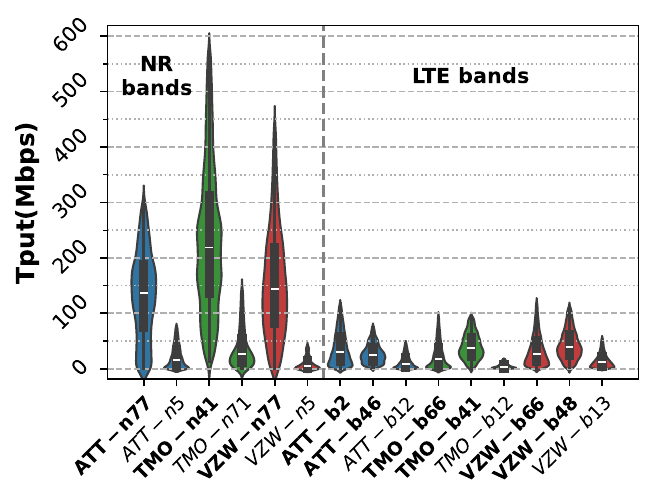}
    \vspace{-2.5em}
    \caption{DL throughput comparison of NR and LTE in low-bands (normal) and mid-bands (bolded)}
    \label{fig:tput}
    \vspace{-1.5em}
\end{figure}

\begin{figure}[t]
    \includegraphics[width=\linewidth]{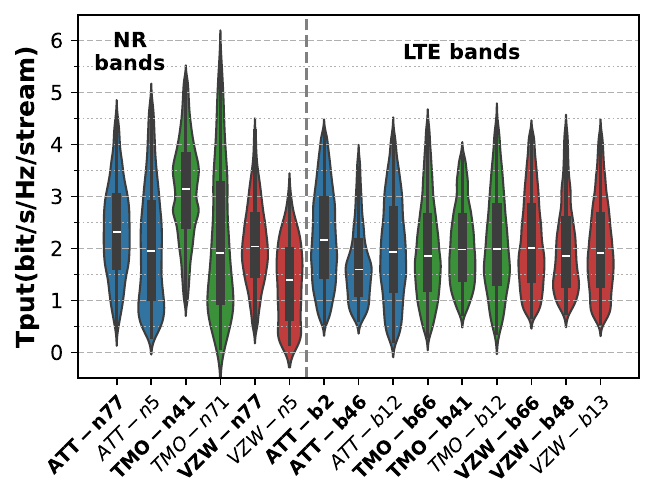}
    \vspace{-2.5em}
    \caption{Normalized DL throughput comparison of NR and LTE in low-bands (normal) and mid-bands (bolded)}
    \label{fig:tput-norm}
    \vspace{-1.5em}
\end{figure}

\subsection{Comparison of Throughput Performance Between Low- and Mid-Band Channels}\label{sec:tput-comp}

\noindent\textbf{\textit{Downlink throughput:}}
Fig.~\ref{fig:tput} shows the comparison of downlink physical layer throughput between low- and mid-band LTE and NR, as reported by XCAL during all of our driving measurements. 
We observe a considerably higher downlink throughput on the mid-band NR channels compared to the low-band counterparts. As discussed in a prior section, this can be explained by the wider bandwidths. The highest median throughput of 218 Mbps is attained by TMO-n41 with a combination of two NR channels of 40 and 100 MHz bandwidth.
For the LTE bands, we similarly observe higher throughputs on mid-band channels compared to the low-bands. The highest median throughput of 31 Mbps is achieved by ATT-b2 with 20 MHz bandwidth. Since the channels' block error rate (BLER) is similar (omitted for brevity), we conclude that the increase in median throughput in NR is due to the wider bandwidth.

For deeper analysis, we examine channel spectral efficiency. This involves normalizing the throughput of each channel by its bandwidth and number of MIMO layers, allowing us to directly compare how effectively each channel utilizes its allocated spectrum. We define normalized throughput: $Tput_{norm} = Tput_{bps} / (N_{RB} * SCS_{Hz} * 12) / N_{layer}$, where $Tput_{bps}$ is the throughput in bits/second, $N_{RB}$ is the average number of resource blocks (RBs) allocated to the UE over one second, $SCS_{Hz}$ is the subcarrier spacing (SCS) in Hz, and $N_{layer}$ is the number of MIMO layers used.
To determine the instantaneous bandwidth usage, we multiply $N_{RB}$ by $SCS_{Hz}$ and 12, given that there are 12 subcarriers in each RB.
Since we use RB to normalize throughput, the difference between TDD configurations in the mid-band channels should not make a difference. However, we observe that all operators in the mid-band channels use the same TDD configuration of 7 slots for DL and 2 slots for UL, with a slot length of 0.5 ms.
Note that when normalizing LTE throughput, RI is used due to the lack of the MIMO layer number KPI in LTE. This is viable since we observe a Pearson correlation of 0.95 between the RI and MIMO layers in our NR data, which is expected since RI is a part of Channel State Information (CSI) feedback to decide the number of MIMO layers.

As computed above, Fig.~\ref{fig:tput-norm} compares the normalized throughput (spectral efficiency) of NR and LTE channels. VZW-n5 exhibits the lowest median throughput, while TMO-n41 achieves the highest (3.14 bit/s/Hz/stream). Other channels fall between 1.9 and 2.3~bit/s/Hz/stream, analogous to the theoretical capacity of uncoded QPSK of 2~bit/s/Hz. Since we previously observed a higher throughput from channels with higher bandwidth, this strongly indicates that the increase in throughput from LTE to NR can be attributed primarily to the wider bandwidth and number of MIMO layers. The exception being TMO-n41, which has a much higher spectral efficiency due to its dense deployment (compared to both ATT and VZW) and larger number of beams (compared to VZW), both of which lead to improved overall signal strength and hence spectral efficiency. Furthermore, the stark contrast with TMO-b41, an LTE channel in the same frequency band, confirms that TMO-n41's superior performance stems from its denser deployment. Finally, we observe a lower normalized throughput in ATT-b46 compared to VZW-b48, where both are shared frequencies.
In the next sections, we will further analyze the contribution of RSRP, modulation, and number of MIMO layers to the normalized throughput.

\begin{figure}[t]
    \vspace{-.5em}
    \includegraphics[width=\linewidth]{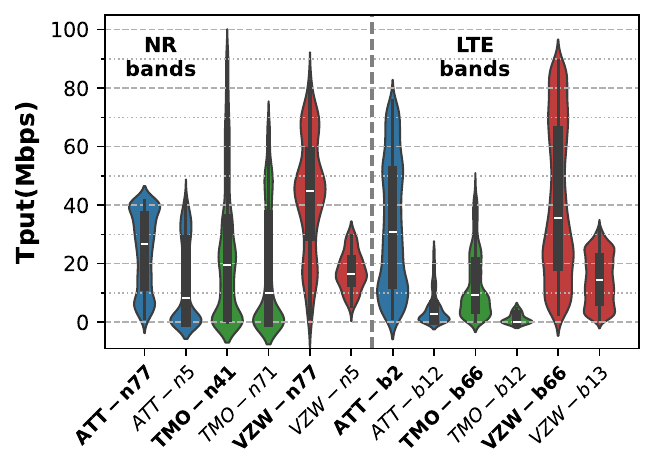}
    \vspace{-2.5em}
    \caption{UL throughput comparison of NR and LTE in low-bands (normal) and mid-bands (bolded)}
    \label{fig:ultput}
    \vspace{-1.5em}
\end{figure}

\noindent\textbf{\textit{Uplink throughput:}}
We omit the complete uplink throughput analysis due to lack of space and limited new insights. Firstly, no uplink channel aggregation was observed within either LTE or NR deployments, even though aggregation is possible between one LTE and one NR channel in the NR-NSA deployment.
Secondly, our data lacked information on the number of uplink MIMO layers parameter, and the NR RI did not correlate with the number of NR layers. Assuming LTE uses 1 stream (based on 97\% of NR data using 1 stream), we see in Fig.~\ref{fig:ultput} that overall uplink throughputs are lower than their downlink counterparts for all representative bands due to the fewer streams. All mid-band channels perform better than the low-band counterparts due to higher bandwidth, while all NR channels show marginal improvements over their LTE counterparts. We did not observe ATT-b46, TMO-b41, and VZW-b48 utilized for uplink transmissions.
Lastly, we observe a higher normalized uplink throughput (median 3.4~bit/s/Hz/stream over all representative LTE and NR channels) compared to downlink. The lower number of uplink layers likely improves stream robustness against fading and errors.


\begin{figure}[t]
    \vspace{-.5em}
    \begin{subfigure}{.98\linewidth}
    \includegraphics[width=\linewidth]{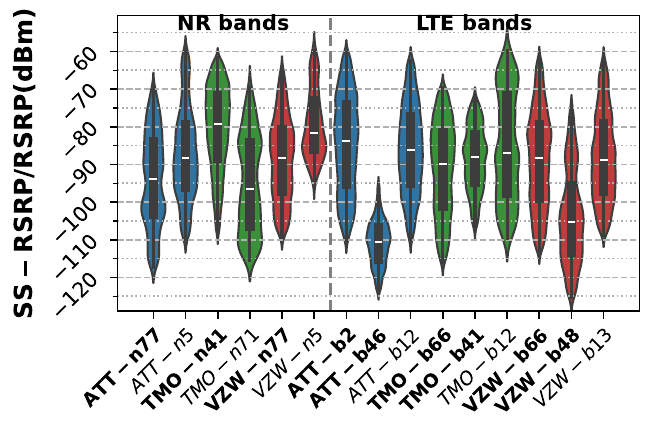}
    \vspace{-2em}
    \caption{CDF of NR SS-RSRP and LTE RSRP}
    \label{fig:rsrp}
    \end{subfigure}
    \\
    \begin{subfigure}{.98\linewidth}
    \includegraphics[width=\linewidth]{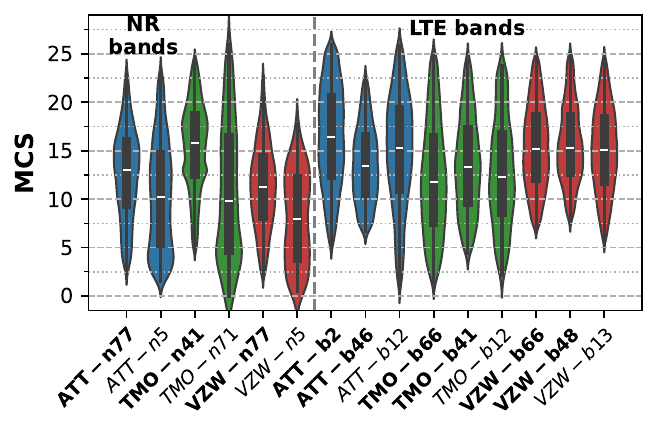}
    \vspace{-2em}
    \caption{CDF of NR and LTE downlink MCS}
    \label{fig:mcs}
    \end{subfigure}
    \\
    \begin{subfigure}{.98\linewidth}
    \includegraphics[width=\linewidth]{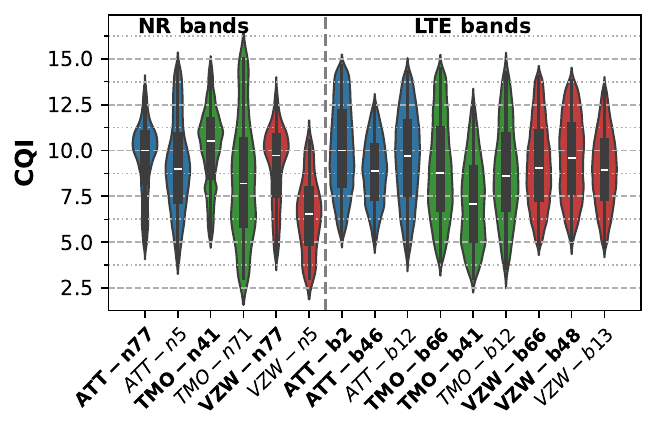}
    \vspace{-2em}
    \caption{CDF of NR and LTE CQI}
    \label{fig:cqi}
    \end{subfigure}
    \caption{Comparison of RSRP, MCS, and CQI of NR and LTE in low-bands (normal) and mid-bands (bolded).}
    \label{fig:rsrp-mcs-cqi}
    \vspace{-2em}
\end{figure}


\subsection{Normalized DL Throughput Impact Factor Analysis}\label{sec:contrib-analysis}

\noindent\textit{\textbf{Contribution of RSRP, MCS, and CQI to normalized DL throughput:}}
Fig.~\ref{fig:rsrp} shows a comparative analysis of the Synchronization Signal RSRP (SS-RSRP) on NR channels, and RSRP on LTE channels.
Both ATT and VZW exhibit higher SS-RSRP values on their low-bands in comparison to their mid-band counterparts: this is due to better propagation characteristics of the low-bands. However, mid-band TMO-n41 displays higher SS-RSRP (median of -79 dBm) compared to its low-band counterpart, n71, which indicates a denser NR deployment to overcome the propagation loss at the mid-bands. Furthermore, TMO consistently displays RSRP values $\sim$12 dB higher than other operators in all of our NR data: this is due to a combination of dense deployment, multiple beams/PCI, and lower frequency.
In LTE, we observe the similarity of RSRP distribution between the channels. The highest median RSRP of -83 dBm is achieved by ATT-b2, which is reflected in the normalized throughput: the highest median downlink throughput of 2.18 bit/s/Hz/stream over all LTE channels. On the other hand, ATT-b46 and VZW-b48 show lower RSRP due to the limitations of transmit power.
Between NR and LTE, TMO-n41 stands out due to its denser deployment and higher number of beams.  

Fig.~\ref{fig:mcs} shows the distribution of allocated downlink MCS, which correlates well with the distribution of normalized downlink throughput in LTE and NR.
For instance, the best and worst median MCS in NR are achieved by TMO-n41 and VZW-n5, respectively, which correspond to the best and worst median normalized downlink throughput. This is expected since higher MCS delivers higher spectral efficiency but can only be used in good signal conditions.
Further, we found RRC messages (``pdsch-Config $>$ mcs-Table := qam256'' in NR and ``cqi-ReportConfig $>$ altCQI-Table-r12 := all subframes'' in LTE) which indicate
that both LTE and NR used the MCS Table 2 as described in Table 5.1.3.1-2 of \cite{3gpp17-38214}, making comparison feasible. We observe a lower median of MCS in NR channels compared to LTE, except for TMO-n41, again due to the excellent channel condition guaranteed by its denser deployment.

We further compare the CQI, which indicates channel conditions from the UE's perspective.
Fig.~\ref{fig:cqi} illustrates the comparison of CQI between the representative LTE and NR channels. Since the BS uses CQI to decide the MCS selection, the distribution of CQI aligns with its respective MCS values: the highest median CQI is attained by TMO-n41, similar to its median MCS, and vice versa with VZW-n5.
On LTE, we also observe a similar distribution of LTE CQI with MCS. This further shows that normalized throughput is mainly influenced by the overall channel condition reported by UE, rather than just RSRP.
However, since we cannot ascertain whether the CQI table used by NR and LTE networks are the same, we cannot make a direct comparison between them.


\begin{figure}[ht]
    \vspace{-1em}
    \includegraphics[width=\linewidth]{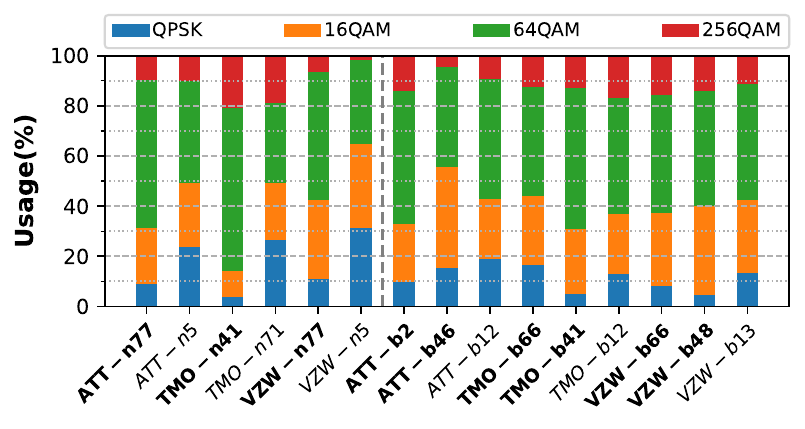}
    \vspace{-2em}
    \caption{Proportional usage of modulation modes of NR and LTE in low-bands (normal) and mid-bands (bolded).}
    \label{fig:modulation}
    \vspace{-.5em}
\end{figure}

\noindent\textit{\textbf{Comparison of DL modulation modes:}}
Fig.~\ref{fig:modulation} shows the modulation modes used for the 12 representative low- and mid-band channels as defined in a previous section. Only modulation modes from QPSK to 256-QAM are observed in our campaign. TMO-n41 shows higher usage of 64-QAM and 256-QAM modulation, corresponding to the channel's high normalized downlink throughput. 
Conversely, the VZW-n5 channel uses a higher proportion of QPSK and 16-QAM modes, which also explains the low normalized throughput of this channel.
For LTE, ATT-b2 mainly uses a combination of 64-QAM and 256-QAM modes, and this results in the highest normalized downlink throughput performance among the LTE channels. Moreover, the resemblance in modulation usage among the other LTE channels corresponds to the similarity in their normalized throughput.

Comparing the modulation modes between the NR and LTE channels, we do not observe a significant improvement, \ie no indication that higher modulation is more available in NR compared to LTE, except for the TMO NR channels n41 and n71, which use 256-QAM more often than the other carriers. This is due to the fact that TMO exhibits better signal conditions in general, which confirms that spectral efficiency improvements are only possible if the overall signal strength improves, through a combination of dense deployments and usage of more beams.

\begin{figure*}[t]
    \begin{subfigure}{.32\textwidth}
    \includegraphics[width=\linewidth]{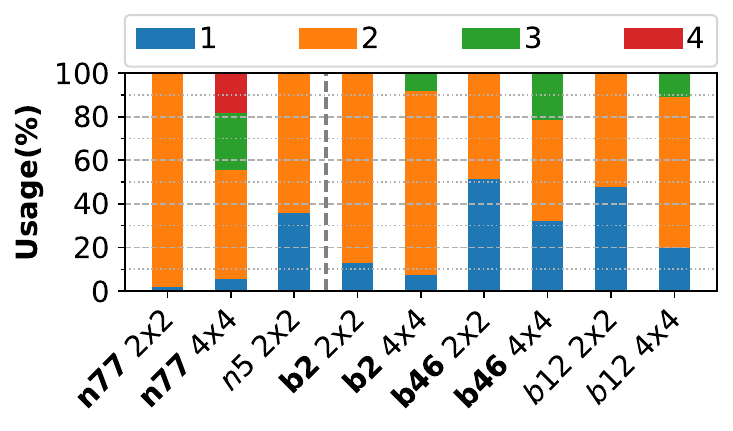}
    \vspace{-2em}
    \caption{ATT}
    \label{fig:cdf-mimo-att}
    \end{subfigure}
    \hfill
    \begin{subfigure}{.32\textwidth}
    \includegraphics[width=\linewidth]{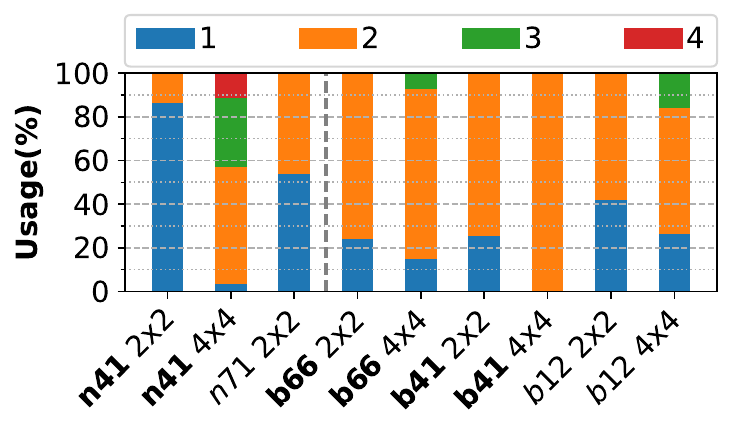}
    \vspace{-2em}
    \caption{TMO}
    \label{fig:cdf-mimo-tmo}
    \end{subfigure}
    \hfill
    \begin{subfigure}{.32\textwidth}
    \includegraphics[width=\linewidth]{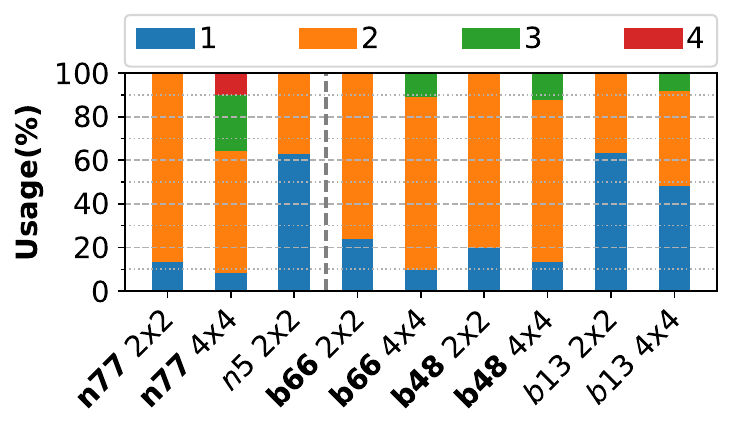}
    \vspace{-2em}
    \caption{VZW}
    \label{fig:cdf-mimo-vzw}
    \end{subfigure}
    \vspace{-.5em}
    \caption{RI value for the operators, channels, and MIMO modes.}
    \label{fig:cdf-mimo-nr-lte}
    \vspace{-1.5em}
\end{figure*}


\subsection{Comparison of MIMO Performance in NR and LTE}\label{sec:mimo-comp}

\noindent\textit{\textbf{Analysis of MIMO performance in terms of RI and MIMO modes:}}
In both LTE and NR, RI is the MIMO channel rank as calculated by the UE and transmitted back to the BS for MIMO layer decision. Fig.~\ref{fig:cdf-mimo-nr-lte} shows the comparison of RI for the three operators on the representative low- and mid-band channels selected in a previous section. The data for each channel is categorized by MIMO modes (\ie 2x2, 4x4) as reported by XCAL. First, we observe that the NR low-band channels on all operators do not utilize 4x4 MIMO modes, while the LTE counterparts do.
Fig.~\ref{fig:cdf-mimo-att} compares the RI value between NR and LTE channels of ATT. We observe an RI value of 2 being the most common in all channels, even when 4x4 MIMO mode is available. For instance, less than 20\% of data on ATT-n77 reported an RI of 4 even when the 4x4 mode is used, while RI 4 is not seen in the corresponding b2 and b12 channels.
Fig.~\ref{fig:cdf-mimo-tmo} similarly shows an RI value of 2 being most common for TMO, with the exception of n41 and n71 using 2x2 mode, where the majority of the data had an RI of 1.
Fig.~\ref{fig:cdf-mimo-vzw} also demonstrates an RI value of 2 being most used for VZW, except for n5 with 2x2 and b13 with both 2x2 and 4x4 modes.

This result is very significant since it indicates that even though higher-order MIMO modes may be implemented, the physical channel rank may not support all available MIMO layers. This suggests that increasing MIMO order in future generations may not be the best way to improve throughput in the real-world. In fact, TMO-n41 has the best throughput performance, but it achieves this with fewer MIMO layers on average compared to ATT-n77 and VZW-n77, which are all mid-band NR channels.

\noindent\textit{\textbf{Analysis of MU-MIMO in NR:}}
While operators claim to have implemented 5G MU-MIMO in test settings~\cite{tmo-mumimo,vzw-mumimo}, its widespread deployment remains uncertain.
To address this question, we utilize the Precoding Matrix Indicator (PMI) which is a part of the CSI feedback to the BS. Moreover, it conveys information about the precoding matrix that should be used in the downlink transmission.
Utilizing XCAL, we collected PMI values from all the UEs and observed PMI with indices: i(1;1), i(1;2), i(1;3), and i(2). This indicates the usage of Type 1-Single Panel codebook with 2--4 MIMO layers~\cite{3gpp17-38214}.
Further, we analyzed the RRC messages captured in all of our UEs and found ``mimo-Parameters $>$ codebookParameters $>$ typeI $>$ singlePanel $>$ mode := mode1'' message which further indicates the usage of the above codebook. 
Type 1 codebooks are used for Single User-MIMO (SU-MIMO) and only use a single beam to calculate CSI feedback \cite{qin2023review}.
Lastly, we did a stationary experiment in Minneapolis in November 2023, where we initiated downlink traffic with up to 4 UEs on the same operator, observing their exact RB allocation in the radio frame. Our hypothesis is that when MU-MIMO is enabled, at least one RB will be allocated in the same slot to two different UEs. We did not observe this in our data.

\begin{figure}[ht]
    \centering
    \includegraphics[width=\linewidth]{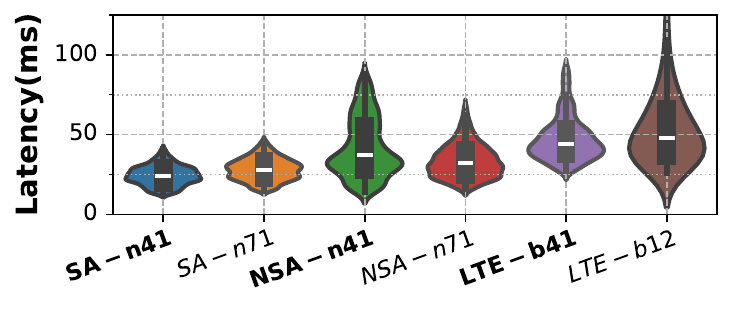}
    \vspace{-2em}
    \caption{Comparison of latency performance in TMO low-bands (normal) and mid-bands (bolded)}
    \label{fig:ping}
    \vspace{-1.3em}
\end{figure}

\begin{figure}[ht]
    \centering
    \includegraphics[width=\linewidth]{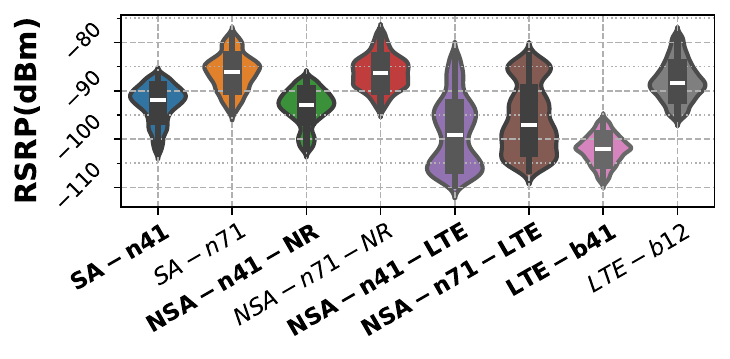}
    \vspace{-2em}
    \caption{Comparison of RSRP in TMO latency measurements}
    \label{fig:ping-rsrp}
    \vspace{-1.3em}
\end{figure}

\subsection{Comparison of Latency Performance in TMO Networks}\label{sec:lat-comp}

We conducted a focused latency measurement in Minneapolis during March 2024, comparing the LTE and NR performance. Specifically, we focus on TMO, which is the only operator that has deployed NR in both SA and NSA modes. Similar to our throughput measurement campaigns, we collected data while driving, using 6$\times$S22+. Using XCAL, we collected signal parameters and round-trip latency results using the included ping tool. We defined two ping targets: Google's cloud DNS server (8.8.8.8) and AWS Local Zone (server located in the same city as the end-user)~\cite{awslocalzone}. Additionally, we utilized XCAL to limit the phones to the following networks and bands: SA-n71, SA-n41, NSA-n71, NSA-n41, LTE-b14 and LTE-b41. This resulted in 12 distinct data categories, each containing an average of 1,774 data points (\ie $\sim$30 minutes of data collection per category).

We observe higher latency to Google DNS compared to Local Zone, which is an expected behavior. Hence, Fig.~\ref{fig:ping} shows the combined latency results (ping to Google and Local Zone) across all bands for brevity. Among the network types, SA bands exhibit the best latency performance, likely due to their simpler architecture. NSA bands, which combine NR and LTE, achieve the next best results. Fig.~\ref{fig:ping-rsrp} also combines the RSRP measured for the Google and Local Zone measurements. Since both NSA-n71 and NSA-n41 utilize either b2 or b66 as the anchor LTE band, we separate those categories with "LTE" or "NR" suffixes to indicate if the RSRP is from the anchor LTE or the secondary NR channel. As expected, we observe higher RSRP on lower-band channels. Interestingly, the LTE anchor channels in NSA bands exhibit a wider spread of RSRP values (-110 to -80~dBm). This lower and more variable RSRP in the LTE channels, combined with the overhead of using both NR and LTE, likely contributes to the slightly higher latency of NSA compared to SA bands. However, both SA and NSA offer significant latency improvements over traditional LTE.

\section{Conclusions and Future Work}
\label{sec:conclusions}

We have presented a comprehensive, real-world evaluation of the performance of low- and mid-band 4G and 5G networks, examining the contribution of various system parameters---such as SSB index, RSRP, modulation, and MIMO---to throughput and latency, and presented conclusions on the reasons for the improvement in 5G throughput over 4G for all three US operators.
It is clear from the data and analyses that the high downlink throughput achieved by mid-band NR channels can be attributed primarily to the higher channel bandwidth and improved signal strengths. When normalized over bandwidth and number of MIMO streams, there is only a marginal improvement in throughput over NR, except for TMO in n41 where we show that the dense deployment and larger number of beams used deliver higher signal strength on average, leading to improved spectral efficiency through the use of higher modulation modes such as 256 QAM.
VZW in n77 uses fewer beams and less dense deployments thus leading to lower normalized throughput compared to other operators.
On the other hand, the 5G uplink performance only shows marginal improvement over 4G due to the lower number of MIMO streams.
Regarding MIMO performance, we observed a marginal increase in the usage of the full capability 4x4 MIMO with 4 layers in NR: this indicates that increasing MIMO modes beyond 4x4 may not be the right approach to increasing throughput since the physical channel may not have sufficient diversity to support more than 4 layers.
The deployment of MU-MIMO was not observed in our campaign, which is confirmed through RRC messages and a focused experiment with multiple UEs.
Lastly, we observed significant 5G latency improvements over 4G on TMO, with the SA configuration delivering the best performance due to its simpler architecture and TMO's denser deployment strategy.

The cellular industry will continue to innovate and deploy new cellular technologies, with governmental agencies opening new spectrum for wireless uses. Recently, the US National Telecommunications and Information Administration (NTIA) announced the intent to study the lower 3 GHz (3.1-3.45 GHz), the 7 GHz band, and other bands for future wireless broadband uses~\cite{ntia2023strategy}. These new spectrum bands will further increase 5G/NextG capabilities by offering increased bandwidth. However, operators and device manufacturers should also aim to implement and optimize new 5G features such as MU-MIMO and higher modulation, which require better signal conditions that can be delivered by dense deployments and increasing the number of beams used. We plan to continue investigating the roll-out of 5G improvements: by quantifying the improvement of various features, such analyses are crucial in informing the development of future standards and deployment architectures.
We provided public access to our data at \url{https://dx.doi.org/10.21227/64wp-sy79}.



\bibliographystyle{IEEEtran}
\bibliography{main}

\begin{thebibliography}{10}
\providecommand{\url}[1]{#1}
\csname url@samestyle\endcsname
\providecommand{\newblock}{\relax}
\providecommand{\bibinfo}[2]{#2}
\providecommand{\BIBentrySTDinterwordspacing}{\spaceskip=0pt\relax}
\providecommand{\BIBentryALTinterwordstretchfactor}{4}
\providecommand{\BIBentryALTinterwordspacing}{\spaceskip=\fontdimen2\font plus
\BIBentryALTinterwordstretchfactor\fontdimen3\font minus \fontdimen4\font\relax}
\providecommand{\BIBforeignlanguage}[2]{{%
\expandafter\ifx\csname l@#1\endcsname\relax
\typeout{** WARNING: IEEEtran.bst: No hyphenation pattern has been}%
\typeout{** loaded for the language `#1'. Using the pattern for}%
\typeout{** the default language instead.}%
\else
\language=\csname l@#1\endcsname
\fi
#2}}
\providecommand{\BIBdecl}{\relax}
\BIBdecl

\bibitem{carpenter2023multi}
J.~Carpenter, W.~Ye, F.~Qian, and Z.-L. Zhang, ``Multi-modal vehicle data delivery via commercial {5G} mobile networks: An initial study,'' in \emph{2023 IEEE 43rd International Conference on Distributed Computing Systems Workshops (ICDCSW)}.\hskip 1em plus 0.5em minus 0.4em\relax IEEE, 2023, pp. 157--162.

\bibitem{bogale2016massive}
T.~E. Bogale and L.~B. Le, ``Massive {MIMO} and {mmWave} for {5G} wireless {HetNet}: Potential benefits and challenges,'' \emph{IEEE Vehicular Technology Magazine}, vol.~11, no.~1, pp. 64--75, 2016.

\bibitem{uwaechia2020comprehensive}
A.~N. Uwaechia and N.~M. Mahyuddin, ``A comprehensive survey on millimeter wave communications for {Fifth-Generation Wireless Networks}: {F}easibility and challenges,'' \emph{IEEE Access}, vol.~8, pp. 62\,367--62\,414, 2020.

\bibitem{3gpp17-38214}
{3rd Generation Partnership Project}, ``{TS} 38.214 {5G}; {NR}; {P}hysical layer procedures for data version 17.1.0 {R}elease 17,'' 2022.

\bibitem{midband}
Nokia, ``{5G} spectrum bands explained,'' \url{https://www.nokia.com/thought-leadership/articles/spectrum-bands-5g-world/}, accessed: 2023-11-26.

\bibitem{tmb-2.5GHz}
\BIBentryALTinterwordspacing
M.~Dano, ``The quiet brilliance of {T}-{M}obile's {5G} spectrum strategy,'' 2022, accessed: 2023-11-26. [Online]. Available: \url{https://www.lightreading.com/5g/the-quiet-brilliance-of-t-mobile-s-5g-spectrum-strategy}
\BIBentrySTDinterwordspacing

\bibitem{att-3.7GHz}
\BIBentryALTinterwordspacing
M.~Alleven, ``{AT\&T} takes advantage of early {C}-band clearing,'' 2023, accessed: 2023-11-26. [Online]. Available: \url{https://www.fiercewireless.com/5g/att-takes-advantage-early-c-band-clearing}
\BIBentrySTDinterwordspacing

\bibitem{vzw-3.7GHz}
\BIBentryALTinterwordspacing
K.~Schulz, ``Verizon turbo charges its {5G} network with the addition of more spectrum,'' 2023, accessed: 2023-11-26. [Online]. Available: \url{https://www.verizon.com/about/news/verizon-5g-network-addition-more-spectrum}
\BIBentrySTDinterwordspacing

\bibitem{li2020beam}
Y.-N.~R. Li, B.~Gao, X.~Zhang, and K.~Huang, ``Beam management in millimeter-wave communications for {5G} and beyond,'' \emph{IEEE Access}, vol.~8, pp. 13\,282--13\,293, 2020.

\bibitem{narayanan2022comparative}
A.~Narayanan, M.~I. Rochman, A.~Hassan, B.~S. Firmansyah, V.~Sathya, M.~Ghosh, F.~Qian, and Z.-L. Zhang, ``A comparative measurement study of commercial {5G mmWave} deployments,'' in \emph{IEEE INFOCOM 2022-IEEE Conference on Computer Communications}, 2022, pp. 800--809.

\bibitem{elijah2015comprehensive}
O.~Elijah, C.~Y. Leow, T.~A. Rahman, S.~Nunoo, and S.~Z. Iliya, ``A comprehensive survey of pilot contamination in massive {MIMO—5G} system,'' \emph{IEEE Communications Surveys \& Tutorials}, vol.~18, no.~2, pp. 905--923, 2015.

\bibitem{jose2011pilot}
J.~Jose, A.~Ashikhmin, T.~L. Marzetta, and S.~Vishwanath, ``Pilot contamination and precoding in multi-cell {TDD} systems,'' \emph{IEEE Transactions on Wireless Communications}, vol.~10, no.~8, pp. 2640--2651, 2011.

\bibitem{lopez2022survey}
D.~L{\'o}pez-P{\'e}rez, A.~De~Domenico, N.~Piovesan, G.~Xinli, H.~Bao, S.~Qitao, and M.~Debbah, ``A survey on {5G} radio access network energy efficiency: Massive {MIMO}, lean carrier design, sleep modes, and machine learning,'' \emph{IEEE Communications Surveys \& Tutorials}, vol.~24, no.~1, pp. 653--697, 2022.

\bibitem{3gpp17-38101-1}
{3rd Generation Partnership Project}, ``{TS} 38.101-1 {5G}; {NR}; user equipment ({UE}) radio transmission and reception; {P}art 1: {R}ange 1 standalone 17.5.0 {R}elease 17,'' 2022.

\bibitem{rochman2023comprehensive}
M.~I. Rochman, V.~Sathya, D.~Fernandez, N.~Nunez, A.~S. Ibrahim, W.~Payne, and M.~Ghosh, ``A comprehensive analysis of the coverage and performance of {4G} and {5G} deployments,'' \emph{Computer Networks}, vol. 237, p. 110060, 2023.

\bibitem{vzw-CBRS}
B.~Fletcher, ``{CBRS} boosts {V}erizon 4g speeds 79\% but power levels show limits,'' \url{https://www.fiercewireless.com/tech/cbrs-boosts-verizon-4g-speeds-79-power-levels-show-limits}, accessed: 2023-11-26.

\bibitem{vzw-att-LAA}
V.~Sathya, M.~I. Rochman, and M.~Ghosh, ``Measurement-based coexistence studies of {LAA} \& {Wi-Fi} deployments in {C}hicago,'' \emph{IEEE Wireless Communications}, vol.~28, no.~1, pp. 136--143, 2020.

\bibitem{3gpp17-38306}
{3rd Generation Partnership Project}, ``{TS} 38.306 {5G}; {NR}; user equipment ({UE}) radio access capabilities 17.0.0 {R}elease 17,'' 2022.

\bibitem{fezeu2023mid}
R.~A. Fezeu, J.~Carpenter, C.~Fiandrino, E.~Ramadan, W.~Ye, J.~Widmer, F.~Qian, and Z.-L. Zhang, ``Mid-band {5G}: A measurement study in {E}urope and {US},'' \emph{arXiv preprint arXiv:2310.11000}, 2023.

\bibitem{narayanan2021variegated}
A.~Narayanan, X.~Zhang, R.~Zhu, A.~Hassan, S.~Jin, X.~Zhu, X.~Zhang, D.~Rybkin, Z.~Yang, Z.~M. Mao \emph{et~al.}, ``A variegated look at {5G} in the wild: performance, power, and {QoE} implications,'' in \emph{Proceedings of the 2021 ACM SIGCOMM 2021 Conference}, 2021, pp. 610--625.

\bibitem{ghoshal2023performance}
M.~Ghoshal, I.~Khan, Z.~J. Kong, P.~Dinh, J.~Meng, Y.~C. Hu, and D.~Koutsonikolas, ``Performance of cellular networks on the wheels,'' in \emph{Proceedings of the 2023 ACM on Internet Measurement Conference}, 2023, pp. 678--695.

\bibitem{liu2023close}
Y.~Liu and C.~Peng, ``A close look at {5G} in the wild: Unrealized potentials and implications,'' in \emph{IEEE International Conference on Computer Communications (INFOCOM’23)}, 2023.

\bibitem{xcal}
\BIBentryALTinterwordspacing
{Accuver}. {XCAL} - {PC} based advanced {5G} network optimization solution. [Online]. Available: \url{https://www.accuver.com/sub/products/view.php?idx=6}
\BIBentrySTDinterwordspacing

\bibitem{iperf}
\BIBentryALTinterwordspacing
{iperf3} - the {TCP}, {UDP} and {SCTP} network bandwidth measurement tool. [Online]. Available: \url{https://iperf.fr/}
\BIBentrySTDinterwordspacing

\bibitem{tmo-mumimo}
M.~Alleven, ``T-{M}obile’s 5{G} network gets capacity boost from {MU-MIMO}: report,'' \url{https://www.fiercewireless.com/tech/t-mobiles-5g-network-gets-capacity-boost-mu-mimo-report}, accessed: 2024-05-09.

\bibitem{vzw-mumimo}
{Signals Research Group}, ``5{G}: The greatest show on earth! vol. 33: {MU-MIMO} and the tower of power,'' \url{https://signalsresearch.com/issue/5g-the-greatest-show-on-earth-33/}, accessed: 2024-05-09.

\bibitem{qin2023review}
Z.~Qin and H.~Yin, ``A review of codebooks for {CSI} feedback in {5G} {N}ew {R}adio and beyond,'' \emph{arXiv preprint arXiv:2302.09222}, 2023.

\bibitem{awslocalzone}
{AWS}, ``{AWS} local zones,'' \url{https://aws.amazon.com/about-aws/global-infrastructure/localzones/}, 2020, accessed: 2024-05-13.

\bibitem{ntia2023strategy}
\BIBentryALTinterwordspacing
{National Telecommunications and Information Administration}. (2023) The national spectrum strategy. [Online]. Available: \url{https://www.ntia.gov/sites/default/files/publications/national_spectrum_strategy_final.pdf}
\BIBentrySTDinterwordspacing

\end{thebibliography}

\end{document}